\documentclass[aps,prb,twocolumn,showpacs,amsmath,amssymb,reprint]{revtex4-2}
\usepackage{dcolumn}
\usepackage{graphicx}
\usepackage{bm}
\usepackage{booktabs}
\begin{document}
	\title{A novel reentrant susceptibility due to vortex and magnetic dipole interaction in a $La_{1.85}Sr_{0.15}CuO_4$  and  $Gd_2O_3$ composite system} 
	\author{Biswajit Dutta$^1$ and A. Banerjee$^2$}
	\affiliation{$^1$Ecole Normale Sup´erieure de Lyon, CNRS, Laboratoire de Physique, F-69342 Lyon, France\\
		$^2$UGC-DAE Consortium for Scientific Research, University Campus, Khandwa Road, Indore-452001, India.}
	
	\begin{abstract}
		
		 A reentrant behavior of temperature dependent magnetic ac-susceptibility (or excess susceptibility
		(ES)) at lower temperature is observed in a composite made of superconductor $La_{1.85}Sr_{0.15}CuO_4$ (LCu) and an insulating paramagnetic salt $Gd_2O_3$ (GdO). The ES exhibits an exponential characteristic that varies with temperature ($\exp,[-\frac{T_0}{T}]$), T0 is characteristics temperature. The characteristics
		temperature,T$_0$, decreases as the effective interface diminishes and the amplitude of the dc magnetic field increases. The creation of ferromagnetic dimers between Gd$^{+3}$ ions in GdO is observed as
		a result of vortex-dipole interaction, which causes the observation of this unusual ES at temperatures much lower than the superconducting onset temperature T$_{S}^{onset}$. This type of ferromagnetic
		dimer formation much below superconducting transition temperature is found comparable with the formation of Yu-Shiba-Rusinov (YSR) state and interaction between these YSR state.
	\end{abstract}
	\maketitle
	\section {Introduction}
	A coherent many-body state of electron pairs with zero spin forms in many materials, leading to the emergence of superconducting (SC) order. This state is associated with an energy gap, which represents the energy required to break a pair by adding or removing an electron, leaving an unpaired electron \cite{r1}. In most materials, interactions that promote magnetism tend to undermine the superconducting order. Local superconducting order depletion can occur when magnetic contaminants are present, and one of the consequence of presence of  magnetic contaminants  is the emergence of an in-gap bound states known as Yu-Shiba-Rusinov (YSR) states \cite{r1,r2,r3,r3,r4,r5,r6,r7,r8,r9}.
	
	The YSR state has been discovered at the edges of a ferromagnet and a superconducting hybrid system \cite{r2,r3,r4,r5}. The localized magnetic moment combines with the superconductor's anti-aligned spin state to create the YSR state, which has discrete spin polarization and an energy $E_b$ that is less than the size of the superconducting energy gap $\Delta$ (i.e., $E_b<\Delta$). This state facilitates a spin triplet ground state [2], and the wave function of this YSR state has a spatially oscillatory, decaying structure \cite{r6,r7,r8,r9}.
	
	When two parallel-spin impurities pair, the hybridization of these YSR states (due to the overlap of the wave functions) results in Shiba bands in these structures [10,11]. Experimental work has focused on examining the Shiba states of single magnetic impurities on superconducting substrates and of coupled dimers of such impurities, with the goal of tailoring the Shiba bands for topological superconductivity \cite{r10,r11,r12,r13,r14,r15,r16,r17,r18,r19}.
		
	Interestingly, an excess susceptibility (ES) or upturn in the susceptibility has been previously reported in a composite system of the superconductor niobium (Nb) with the normal metals silver (Ag) or gold (Au) \cite{r20,r21,r23,r24}. This behavior was initially explained in terms of the formation of a "p-wave" superconducting bound state in the novel metal, due to the proximity of the superconductor. However, the precise cause of this anomalous ES or upturn remains to be unambiguously identified.
	
	Notably, the theoretical work of Roman M. Lutchyn et al. has modeled that the interface between an 's-wave' superconductor and a material with strong spin-orbit coupling, such as the novel metals Ag and Au, can provide a fertile ground for the appearance of Majorana Zero modes \cite{r25}. Furthermore, the zero-energy excitation of the 'Shiba' band is also found to carry information about the topological superconductor and the Majorana modes \cite{r2,r4,r13}.
	
	These proximity-induced effects in hybrid superconducting systems warrant further scrutiny to fully understand the underlying mechanisms behind the observed anomalous magnetic susceptibility behavior. The interplay between superconductivity, magnetism, and spin-orbit coupling at these interfaces presents an intriguing avenue for exploring exotic quantum phenomena and potentially realizing topological superconductivity.
	
	In this work, we report an anomalous excess susceptibility (ES) at very low temperatures in composites of the superconductor $La_{1.85}Sr_{0.15}CuO_4$ (LCu) and the paramagnetic material $Gd_2O_3$ (GdO), which has a large spin magnetic moment of approximately 7$\mu_{b}$. The ES is influenced by changes in the interface between GdO and LCu, and by the application of a dc magnetic field that destroys superconductivity. This ES is similar to phenomena observed in Nb and Au/Ag composite systems.
	
We observed a finite value of the second-order susceptibility ($\arrowvert\chi_2\arrowvert$) even in the absence of a DC magnetic field. The increased susceptibility at low temperatures, along with a nonzero second-order susceptibility ($\arrowvert\chi_2\arrowvert$(T)) at zero-bias DC field, indicates a modulated ferromagnetic or ferrimagnetic interaction between $Gd^{+3}$ spins. The modulation induced by the DC magnetic field suggests that this interaction is mediated through vortices, implying a signature of vortex and magnetic dipole interaction. The scaling of the excess susceptibility with the effective interface between GdO and LCu, along with the magnetic field, suggests the opening of an energy gap with an amplitude smaller than the superconducting energy gap. This phenomenon is further elaborated in the experimental results section. Additionally, these observations indicate the emergence of a spin triplet Yu-Shiba-Rusinov (YSR) type dimer state.
	
To modify the effective interface, we prepared two sets of composites. In the first set, we maintained fixed particle sizes but varied the weight percentages of LCu and GdO. In the second set, we kept the LCu particle size and the weight percentages of LCu and GdO constant, while varying the particle size of GdO. We employed both linear and nonlinear AC susceptibility measurement techniques to study these composites.

It is noteworthy that linear and nonlinear AC susceptibility measurements are crucial tools for characterizing various properties of type-II superconductors, including determining critical thermodynamical parameters such as critical current, critical field, and critical temperature \cite{r26, r27, r28, r29}. Moreover, these techniques are highly effective in distinctly identifying various metastable states such as spin-glass, cluster glass, and superparamagnet \cite{r30, r31, r32, r33, r34, r35, r36}, as well as in studying long-range ordered systems like ferromagnets, ferrimagnets, and antiferromagnets \cite{r37, r38, r39}.
	
	\section{EXPERIMENTAL DETAILS}
	\subsection{Sample preparation and characterization}
	
	\begin{figure}[t]
		\centering
		\includegraphics[width=9.5cm,height=8cm]{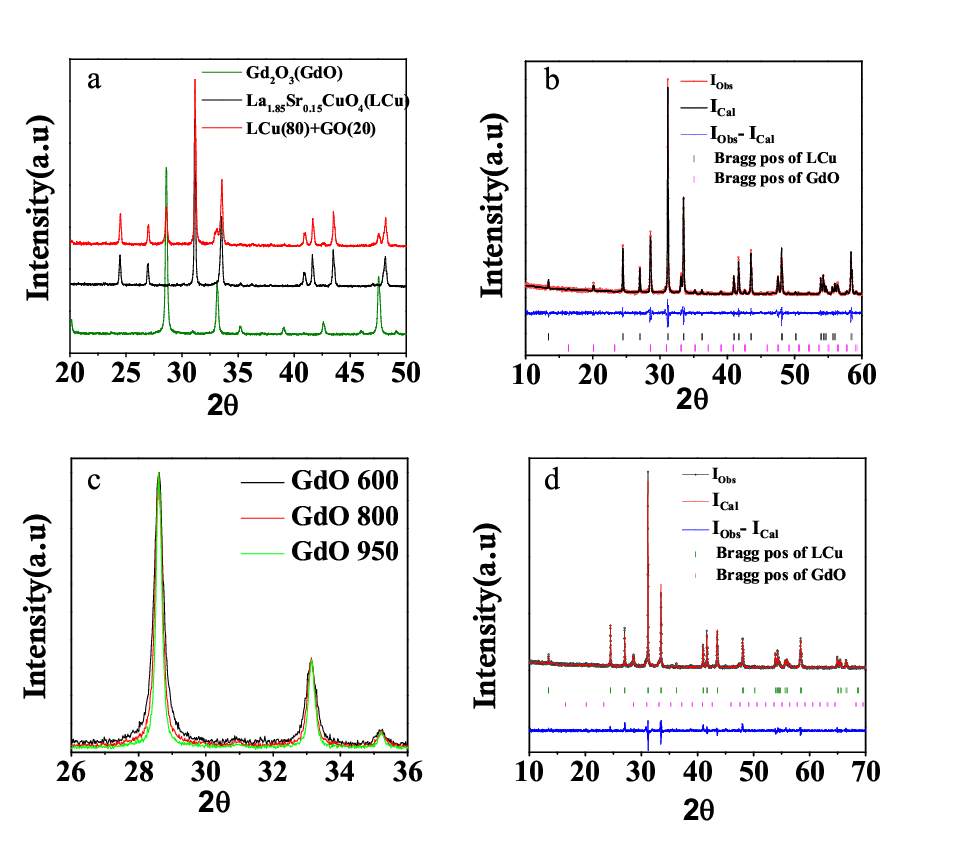}
		\vspace{-5mm}
		\caption{\textit{\small{(Colour Online)(a) Room temperature XRD pattern of composites B3 and the parent the ingredients LCu and GdO respectively.(b) Rielveld refinement of the composite B3(LCu (76\%)+GdO(24\%).(c) XRD pattern of GdO heated at various temperature. (d) Rielveld refinement of the composite B2(LCu (90\%)+GdO600(10\%).}}}
		\label{fig:Fig3.6}
	\end{figure}
	$La_{1.85}Sr_{0.15}CuO_4$ (LCu) sample is prepared by pyrophoric method \cite{r40} and final heat treatment is done at $950^0$C. Then LCu and commercial $Gd_2O_3$ (GdO) powder of 99.99\% purity are mixed in three different weight ratios and grind them very carefully to mix uniformly. Then the mixed powder is palletized at considerable high pressure $\sim$150 kN and heated at  $830^0$C for 4 hrs to prepare the composite. The three composite thus made have the weight percentage and nomenclature as follows:\\LCu (94\%)+GdO(6\%)-B1,\\LCu (90\%)+GdO(10\%)-B2,\\LCu (76\%)+GdO(24\%)-B3,
	
	To make the second set of three composites, nanoparticles of GdO is prepared by pyrophoric method. Then the precursor powder obtained from pyrophoric	method is heated at three different temperatures, like  $600^0$C,  $800^0$C and  $950^0$C to vary the particle size of GdO. Then the precursor powder of LCu and GdO (different particle sizes) are mixed in 90:10 weight ratio (i.e., 90 weight percentage of LCu and 10 weight percent of GdO) and heated at  $550^0$C to prepare the composites whose details and attributed nomenclature is given\\
	LCu (90\%)+GdO600(10\%)-B4,\\ LCu (90\%)+GdO800(10\%)-B5,\\ LCu (90\%)+GdO950(10\%)-B6\\
	\begin{table}
		\caption{Best fitted parameters, PS- particle size, SG- space group, PF- phase fraction} 
		\begin{tabular}{|c| c| c| c| c| c| c| c|} 
			\hline
			Mateial& PS(nm)& SG& a=b(A$ {^0})$  & c(A$ {^0})$& PF& $R_f$\\ [0.3ex] 
			\hline
			(LCu)  &bulk& I4/mmm &3.777(2) &13.222(9)& NA&1.35\\ 
			
			(GdO) &bulk& I213 &10.818(4) & 10.818(4)& NA&1.44 \\
			
			(GdO600) &25& I213  & 10.820(4) &10.820(4)& NA&1.44 \\
			
			(GdO800) &45& I213 &10.815(4)  &10.815(4)& NA&1.44 \\
			
			(GdO950) &60& I213 &10.813(4) &10.813(4)& NA&1.44 \\
			\hline
			B1(LCu)&bulk&I4/mmm & 3.779(1) &13.223(9)&96&1.47 \\
			
			B1(GdO)&bulk&I213&10.819(6) &10.819(6)&4&1.47 \\
			\hline
			B2(LCu)&bulk&I4/mmm & 3.777(1) &13.220(9)&96&1.47 \\
			
			B2(GdO)&bulk&I213&10.819(6) &10.816(6)&4&1.47 \\
			\hline
			B3/( LCu)&bulk&I4/mmm & 3.779(4) & 13.226(6) &76 & 1.44 \\
			
			B3/( GdO)&bulk&I213 & 10.817(6) &10.817(6)&24&1.44 \\
			\hline
			B4/( LCu)&bulk& I4/mmm & 3.782(3) & 13.223(1)&91&1.8 \\
			
			B4/( GdO600)&25& I213 & 10.816(8) &10.816(3)&9&1.8 \\
			\hline
			B5/( LCu)&bulk& I4/mmm & 3.784(1) & 13.228(8)&92&1.6 \\
			
			B5/( GdO800)&45& I213&10.814(9)&10.814(1)&8&1.6 \\
			\hline
			B6/( LCu)&bulk& I4/mmm&3.784(5)&13.227(7)&92&1.87 \\
			
			B6/( GdO950)&60& I213&10.812(1)&10.812(7)&8&1.87 \\
			\hline
		\end{tabular}
	\end{table}
		\begin{figure*}[t]
		\centering
		\includegraphics[width=16cm,height=13cm]{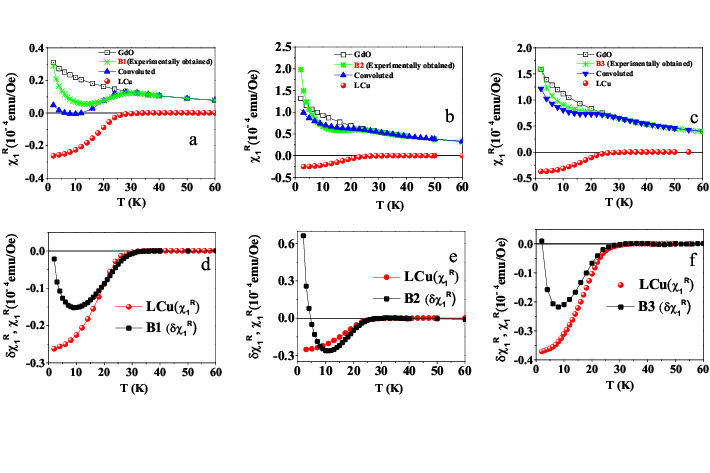}
		\vspace*{-25mm}
		\caption{\textit{\small{(Colour Online) (Colour Online) Ac-susceptibilty measurements are performed in an ac-field of 3 Oe and frequency 231.1 Hz, (a)	Comparative plot of the real part of first order ac-susceptibility ($\chi_1^R$ ) against temperature of the composite B1, LCu, GdO and linear superimposed (or the simple convoluted) $\chi_1^R$ . Similar type of plot for the composite B2 and B3 is shown in (b) and (c),respectively. The temperature dependent deconvoluted susceptibility ($\delta\chi_1^R$ ) plot for B1, B2 and B3 is shown in (d), (e) and (f),respectively, along with them the normalized susceptibility plot ($\chi_{1}^R$) of LCu is also plotted with each deconvoluted graph.}}}
		\label{fig:fig1}
	\end{figure*}
	The samples are characterized by X-ray diffraction (XRD) performed in Bruker X-ray diffractometer from $10^0$- $90^0$ at an interval of $0.02^0$. \texttt{Fig.1a} shows representative XRD patterns of the parent paramagnet (GdO) and superconductor (LCu) along with the two phase Rietveld refinement of the composites. Fig.1b shows the two phase Rietveld refinement of XRD data of the composite B3. \texttt{Fig.1c} shows the XRD pattern of GdO annealed at various temperatures resulting in different particle sizes exemplified by the difference in full width at half maxima (FWHM). Particle sizes obtained from the Williamson Hall Plot are 25\,nm, 45\,nm and 60\,nm for samples annealed at $600^0$C, $800^0$C and $950^0$C respectively. \texttt{Fig.1d} shows the two phase Rietveld refinement of XRD data of the composite B2 i.e. LCu (90\%)+GdO600(10\%).

	From the XRD pattern and the Rietveld refinement of the composite no impurity phase could be identified confirming no chemical reaction taken place between the
	ingredients LCu and GdO. The lattice parameters of LCu and GdO obtained from the two phase Rietveld refinement of all the composites depicts no significant change
	with respect to the corresponding raw materials i.e. LCu and GdO.
	
	The value of the lattice parameters of the corresponding raw materials and the ingredients present in all the composites are provided in Table I. No significant change in the lattice parameters of the ingredients in the composites with respect to their parent or raw material indicates
	that there is no considerable lattice strain is developed in either ingredients while preparing the composites. The details of space group (SG), lattice parameters, phase fraction (PF), particle size(PS) etc. are listed in \texttt{Table\,I}. The PFs obtained from two phase Rietveld refinement are used to calculate the amount of GdO and LCu present in the composite per unit gram (gm) which is used to normalize the ac-susceptibility value of the composite.
	
	\subsection{Magnetic measurements}
	
	The Low field linear and nonlinear magnetic ac-susceptibility measurements have been performed using a homemade ac-susceptometer, which can be operated down to 4.2\,K from 300\,K and the measurements can be done in both cooling and heating cycle with a temperature accuracy of 1\,mK. The estimated sensitivity of the setup is $\sim 10^{-7}$emu [41]. The higher dc-field ($>$ 200 Oe) superimposed ac-susceptibility measurements are performed in MPMS-XL (M/S, Quantum Design). Low field ac-susceptibility measurement probes the spin dynamic at very low field. The magnetization (m) can be expanded with respect to the applied ac-field $h_{ac}$ as
	
	\[ m = m_0+\chi_1 h+ \chi_2 h^2+\chi_3 h^3+\chi_4 h^4....(1)\]
	$\chi_1(\approx\delta m/\delta h$) is linear susceptibility and  $\chi_2$, $\chi_3$, $\chi_4$...  are nonlinear susceptibilities.These nonlinear susceptibilities contain many fruitful information but magnitude of
	these are much smaller (couple of order) than the linear
	susceptibility, therefore they are difficult to measure from
	normal dc-magnetization measurement, but these nonlinear susceptibilities can be easily measured from high sensitive ac-susceptibility measurement \cite{r32,r35,r41}. If the magnetization has an inversion symmetry with respect	to the applied ac-field (h$_{ac}$) then all the even order susceptibilities, like $\chi_2(\approx\delta^2 m/\delta^2 h$), $\chi_4(\approx\delta^4 m/\delta^4 h$) do not appear in the absence of external dc-field (at  h$_{dc}$\,=\,0\,Oe) like paramagnetic and antiferromagnetic materials, but when the inversion symmetry breaks by internal field then $\chi_2$, $\chi_4$.. all the even order susceptibility shows finite value (at h$_{dc}$\,=\,0\,Oe), like in case of ferromagnet, or	ferrimagnetic state \cite{r42,r43}. Even order susceptibilities	also appear when external dc-field is applied, superimposing on ac-field. The third order susceptibility ($\chi_3$) is very useful tool to unambiguously distinguish various metastable states like spin glass (SG), superparamagnet (SPM) etc. [30–36]. $\chi_3$ is also found as very effective tool in determining the universality class of the ferromagnet i.e. to study the nature of magnetic ground state of the corresponding ferromagnet \cite{r37,r38,r39}.
	
	\section{Results and discussions}	
	The combined temperature-dependent plot of $\chi_1$ (the real component of the first-order AC susceptibility), $\chi_1^R(T)$, for the composites B1, B2, and B3 is presented in \texttt{FIG.2}. This plot is accompanied by the $\chi_1^R(T)$ plot of the parent constituents LCu and GdO for comparative purposes. To demonstrate the proximity effect resulting from the interaction between LCu and GdO, the mass-normalized convoluted and deconvoluted components are also included in these graphs. The $\chi_1^R(T)$ plots for the composites (B1, B2, B3 - Green stars), LCu (Red circles), and GdO (Black squares), as well as the convoluted susceptibility of LCu and GdO (Blue triangles), are displayed in \texttt{FIG.2a}. \texttt{FIG.2b} and \texttt{FIG.2c}, respectively.
	
	Convolution is performed by combining the mass-normalized susceptibility data of the parent materials, LCu and GdO. The susceptibility graph of parent GdO is normalized relative to the mass of GdO present in the composite used for susceptibility measurement. Similarly, the susceptibility graph of LCu is normalized relative to the mass of LCu present in the composite. In \texttt{FIG.2a}. \texttt{FIG.2b} and \texttt{FIG.2c}, all graphs converge above the onset temperature of superconductivity (T$_{S}^{(onset)}$). These observations indicate that the magnetization value (spin moment only, as $L=0$ for Gd$^{+3}$ ion) of the mole-normalized GdO present in the composite and the parent GdO (mole-normalized) are equivalent. This suggests that there was no chemical reaction between the LCu and GdO components during the composite preparation.
	
	Since LCu is a Pauli paramagnetic substance, its temperature-dependent susceptibility should be temperature-independent and very small in magnitude above T$_{S}^{(onset)}$ compared to GdO. Thus, it should not affect the Curie-Weiss susceptibility behavior of GdO above T$_{S}^{(onset)}$ \cite{r43}. The amplitude of the experimentally measured susceptibility values of the composites (Green stars) shows a greater magnitude than the convoluted data (Blue triangles) at much lower temperatures (approximately 7 to 9\,K), where GdO starts deviating from Curie-Weiss behavior, as shown in \texttt{FIG.2a}. \texttt{FIG.2b} and \texttt{FIG.2c}. Therefore, this additional susceptibility appears to result from either a reduction in the diamagnetic portion of LCu or the emergence of a specific type of modulated magnetic interaction at the bulk and interface of LCu and GdO.
	
	The diamagnetic fraction of LCu can change for several reasons, including:
	1-Alteration of LCu’s crystal structure and crystalline size during composite preparation.
	2-Chemical reaction between LCu and GdO, which can also destroy superconducting properties.
	3-Pinning of vortices due to magnetic interaction between LCu and GdO across the interface.
	XRD measurements show no significant change in the crystal structure of LCu and GdO during composite preparation and depict the absence of extra impurity peaks other than those of LCu and GdO. Additionally, the same value of $\mu_{eff}$ for the composites and GdO (calculated from Curie-Weiss fitting of the $\chi_1^R(T)$ graph above T$_{S(\text{onset})}$) eliminates the possibilities mentioned in points (1) and (2). The remaining possibility is the interaction between GdO and LCu across the interface, which alters the magnetic properties of the composite relative to the convoluted data.
		\begin{figure*}[t]
		\centering
		\includegraphics[width=23cm,height=10cm]{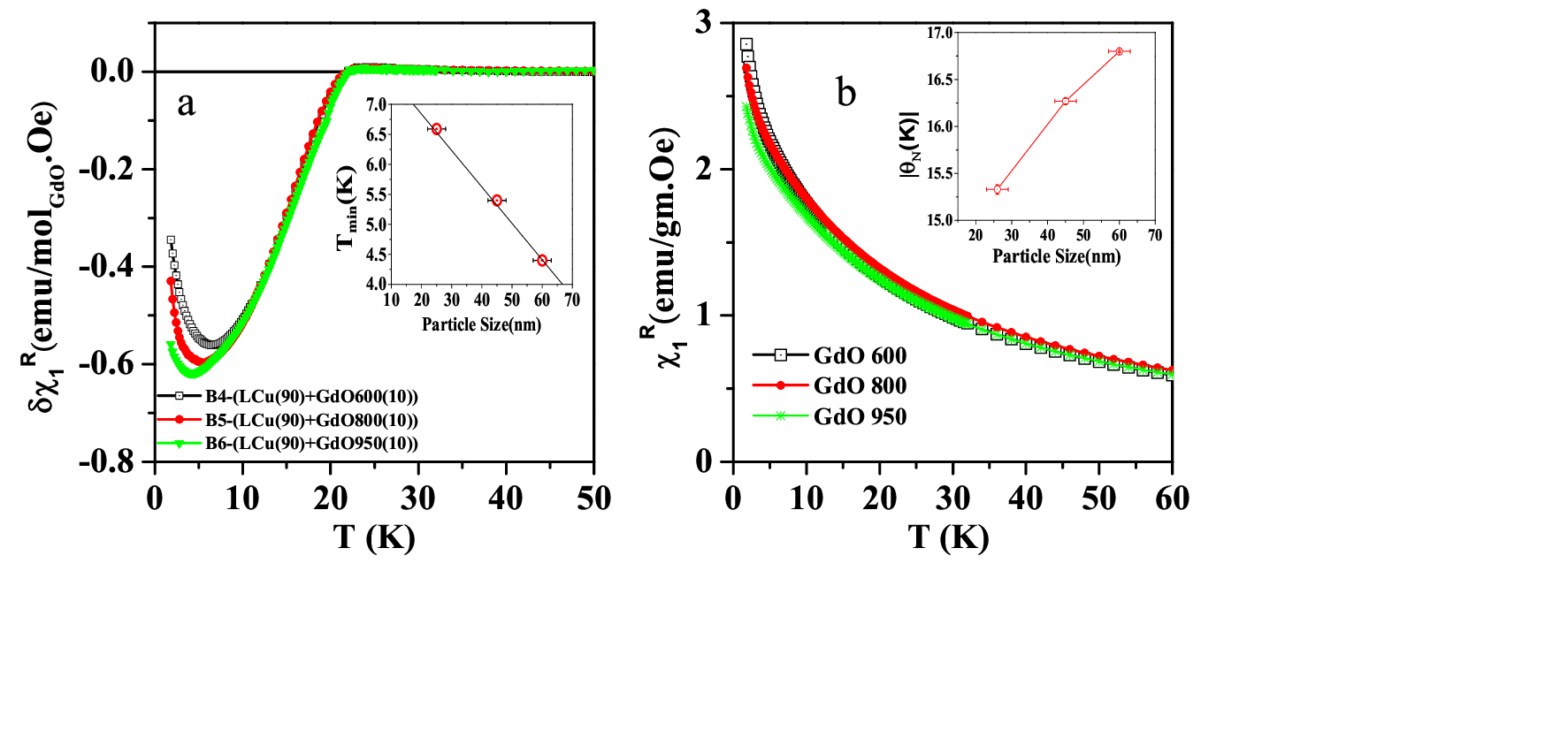}
		\vspace*{-24mm}
		\caption{\textit{\small{(Colour Online) All the measurements has been performed in an ac field of 3 Oe and frequency 231.1 Hz (a) Normalize $\delta\chi_1^R$ against temperature for B4,B5 and B6 composite respectively. Inset shows the plot of $T_{min}$ against the particle size of GdO and the red line corresponds to the straight line fitting. (b) Normalized $\chi_1^R$ is plotted against temperature for GdO600, GdO800 and GdO950 respectively (Inset shows the plot of Curi-Weiss temperature ($\theta_N$) against the particle size of GdO).}}}
		\label{fig:fig2}
	\end{figure*}
	
	\begin{table*}[t]
		\centering	
		\caption{\bf Best fitted parameters} 
		\begin{tabular}{|c|c|c|c|c|}
			\hline 
			Composite & T$_{min}$\,(K) & Paricle size of GdO(nm) & $\arrowvert\theta_{N}\arrowvert$\,(K) & $C_{cal}$(emu/mole.K)  \\ 
			\hline 
			B4 & 6.60 & 25 & 15.3 $\pm$ 0.08 & 8.07$\pm$ 0.05 \\ 
			\hline 
			B5 & 5.40 & 45 & 16.2 $\pm$ 0.04 & 7.90$\pm$ 0.07 \\  
			\hline
			B6 & 4.40 & 60 & 16.8 $\pm$ 0.02 & 7.98$\pm$ 0.05 \\ 
			\hline  
		\end{tabular}\\
	\end{table*}
	To better visualize the changes in magnetic properties due to the mutual interaction between LCu and GdO, a deconvolution operation was performed. Specifically, we subtracted the mass-normalized susceptibility data of GdO from the susceptibility data of the composite. This deconvoluted susceptibility, normalized by the same previously mentioned mass value, is denoted as $\delta\chi_1^R$. The resulting temperature-dependent deconvoluted susceptibility, $\delta\chi_1^R(T)$, is plotted alongside the mass-normalized susceptibility, $\chi_1^R(T)$, of LCu. These combined plots are presented in \texttt{FIG. 2d}, \texttt{FIG. 2e}, and \texttt{FIG. 2f} for samples B1, B2, and B3, respectively.
	
	It is noteworthy that there is no discernible difference between the values of $\delta\chi_1^R(T)$ and $\chi_1^R(T)$ for LCu around the onset temperature, $T_{S}^{(onset)}$. This indicates that the chemical composition of LCu in the composite remains unchanged from that of the parent LCu. However, a sudden change in $\delta\chi_1^R(T)$ below 10\,K ($\sim$\,-10$^{-5}$\,emu to +10$^{-5}$\,emu) suggests that the interaction extends beyond the surface or interface, penetrating into the bulk of one of the components.
	
	In the $\delta\chi_1^R(T)$ plot for B1, a minimum is observed at 7 K, followed by an upturn below this temperature. This behavior is absent in the parent LCu, which does not exhibit any anomalies in this temperature range. The composite B2, as shown in \texttt{FIG. 2e}, exhibits the most significant anomaly in low-temperature $\delta\chi_1^R(T)$ among all the composites. Although the onset temperature, $T_{S}^{(onset)}$, remains constant across all composites, the minimum value of $\delta\chi_1^R(T)$ is observed at different temperatures: 7 K for B1, 11 K for B2, and 8 K for B3. This variability suggests that the anomalous excess susceptibility is an interface-induced phenomenon resulting from the magnetic interaction between GdO and LCu.

	\begin{figure*}[htbp]
		\centering
		\includegraphics[width=16cm,height=7cm]{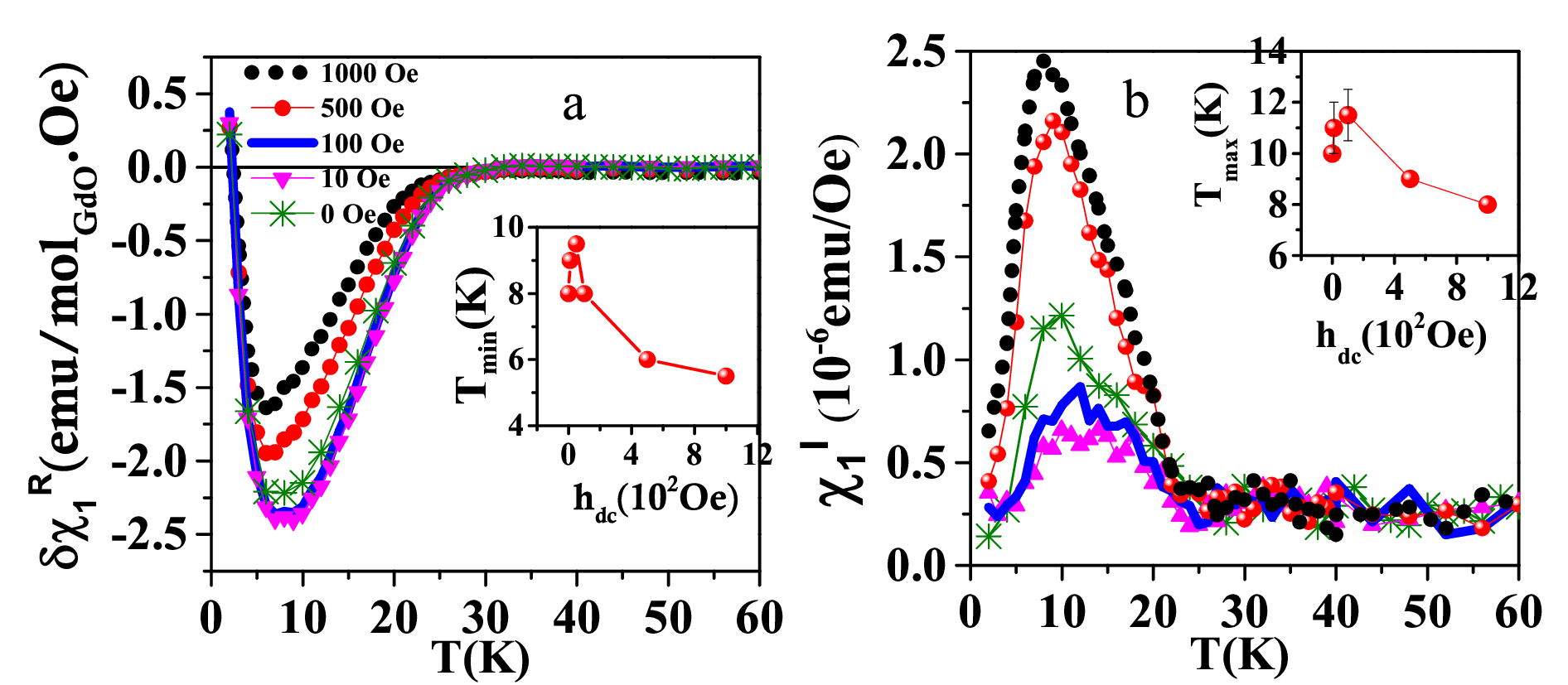}
		\vspace*{-2mm}
		\caption{\textit{\small{(Colour Online) All the measurements are done in an ac field of 3 Oe and frequency 231.1 Hz in B3 composite (a) $\delta\chi_1^R$ is plotted against temperature in superimposed dc field of amplitude 0, 10, 100,500 and 1000 Oe respectively, (Inset shows dc field dependent plot of $T_{min}$). (b)$\chi_1^I$ is plotted against temperature for various amplitude of dc fields (0,10,100,500 and 1000\,Oe). Inset shows the dc field dependent plot of T$_{max}$.}}}
		\label{fig:fig3}
	\end{figure*}
	
	The interface effect is further corroborated by modulating the effective interface between LCu and GdO through varying the GdO particle size while maintaining a constant mass ratio. Herein, we present the results corresponding to the composition (mass ratio) of composite B2 (i.e., LCu(90) + GdO(10)). The effective interface between LCu and GdO is altered with changes in particle size due to the resultant modifications in the surface-to-volume ratio of GdO. The plot of $\delta\chi_1^R$(T) for the composite containing GdO600, GdO800, and GdO950 (denoted as B4, B5, and B6, respectively) is illustrated in \texttt{FIG.3a}. The values of $\delta\chi_1^R$(T) for all these composites are calculated similarly to the method described previously. The results indicate that T$_{S}^{(onset)}$ remains constant across all composites (i.e., B4, B5, and B6). Additionally, the unchanged values of orthorhombic distortion suggest no variation in the hole concentration of LCu, implying the chemical state of LCu in all composites remains consistent.
	
	However, a progressive decrease in T${min}$ with increasing GdO particle size (i.e., for B4, T${min}$ is observed at approximately 6.6 K, and for B6, it is observed around 4.4 K) suggests that the low-temperature anomaly or the existence of electronic scattering (ES) is an interface-induced phenomenon. The specifics of the GdO particle size in composites B4, B5, and B6, along with the Curie-Weiss temperature ($\theta_{N}$(K)), Curie-Weiss constant (C), and T$_{min}$(K) values, are provided in Table II. The inset of \texttt{FIG.\,3a} depicts the variation of T${min}$ with GdO particle size, demonstrating that the interaction between LCu and GdO becomes more pronounced as the GdO particle size decreases.
	
	The effects of particle size reduction include:
	
	Enhanced interaction efficiency between LCu and GdO.
	As shown in \texttt{FIG.\,3b}, the susceptibility value of GdO increases at lower temperatures.	The Neel temperature ($\theta_{N}$) of GdO also decreases, as illustrated in the inset of \texttt{FIG.\,3b}.
	The first result has been previously described, while the second and third observations indicate that the antiferromagnetic (AFM) correlation between the Gd$^{+3}$ spins diminishes with decreasing GdO particle size. This leads to an increase in the number of unpaired Gd$^{+3}$ spins on the surface and within the bulk of GdO \cite{r44, r45, r46}. Due to the heightened interaction probability between GdO and LCu and the larger effective interface, composite B4 exhibits the most significant $\delta\chi_1^R$ anomaly at lower temperatures compared to B5 and B6.
	
	An important observation is that the upturn in $\delta\chi_1^R$ (depicted in \texttt{FIG.\,3a}) is not as sharp as in the $\delta\chi_1^R$(T) graphs of B1, B2, and B3 (shown in \texttt{FIG.\,2d}, \texttt{FIG.\,2e}, and \texttt{FIG.\,2f}). This difference is attributed to the annealing temperatures: B1, B2, and B3 composites are annealed at a higher temperature ($830^0$C), whereas B4, B5, and B6 are annealed at a comparatively lower temperature ($550^0$C). This results in lower compactness (or intergranular contact) in B4, B5, and B6 compared to B1, B2, and B3. The linear decrease in T$_{min}$ and the gradual suppression of $\delta\chi_1^R$(T) below T$_{min}$ with increasing GdO particle size (as shown in the inset of \texttt{FIG.\,3a}) imply that the low-temperature upturn in the $\delta\chi_1^R$(T) graph is an intrinsic mesoscopic effect of the LCu and GdO proximity structure.
	
	To gain a deeper understanding of the observed anomaly, we propose a model wherein the unpaired Gd$^{+3}$ spins interact with vortices via vortex dipole interactions. Given the substantial magnetic moment of the Gd$^{+3}$ spin (7$\mu_B$), it can generate significant local magnetic fields and effectively couple with the vortices within the superconductor. To further investigate this interaction, we conducted dc-field superimposed ac-susceptibility measurements.
	
	The primary function of the applied dc magnetic field is to modulate the vortex dipole interactions. This modulation occurs through the induction of the Lorentz force on the vortices and the Zeeman interaction, which biases the uncompensated Gd$^{+3}$ spins in GdO. The resultant $\delta\chi_1^R$(T) graph is expected to reflect these phenomena. Specifically, the $\delta\chi_1^R$(T) graph at various superimposed dc-fields is depicted in \texttt{FIG.\,4a} (for the B3 composite).
	
	The experimental procedure is as follows:
	
	Measurement: Initial dc-field superimposed ac-susceptibility measurements were performed on each parent component and composite.
	Data Extraction: Subsequently, a deconvolution method was employed to determine the $\delta\chi_1^R$(T) values for each dc-field, consistent with the methodology discussed in the preceding section.
	As illustrated in \texttt{FIG.\,4a}, the diamagnetic fraction and T$_{min}$ of $\delta\chi_1^R$(T) increase for lower dc bias fields (h${dc} <$ 100\,Oe) compared to the $\delta\chi_1^R$(T) at h$_{dc}$ = 0\,Oe. However, for dc bias fields exceeding 100\,Oe, both the diamagnetic fraction and T$_{min}$ decrease. The inset of \texttt{FIG.\,4a} shows the trend of T$_{min}$ versus the superimposed dc-field, indicating an initial increase in T$_{min}$ up to 100\,Oe of dc-field, followed by a subsequent decrease above 100\,Oe. This behavior underscores the complex interplay between the magnetic field, vortex dynamics, and Gd$^{+3}$ spins in the composite.

	\begin{figure*}[t]
		\centering
		\includegraphics[width=16cm,height=7cm]{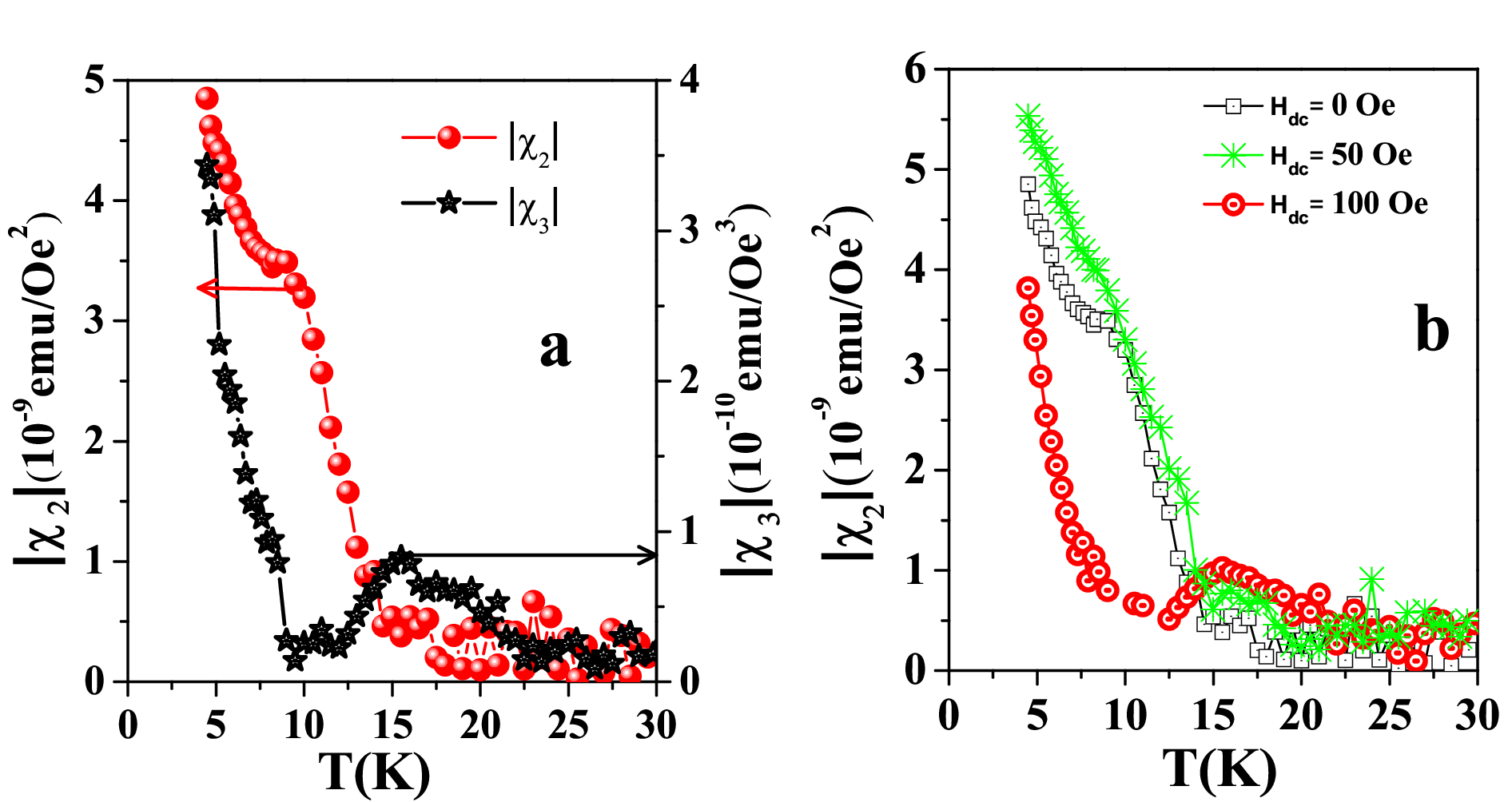}
		\vspace*{-2mm}
		\caption{\textit{\small{(Colour Online) second order ac susceptibility($\arrowvert\chi_2\arrowvert$) is plotted against temperature (Left hand side Y axis is the corresponding scale) and Third order ac susceptibility ($\arrowvert\chi_3\arrowvert$) is plotted against temperature (Right hand side Y axis is the corresponding scale) for composite B3. (b) $\arrowvert\chi_2\arrowvert$ against temperature in three superimposed dc field (0 Oe, 50 Oe and 100 Oe) for composite B3.}}}
		\label{fig:fig4}
	\end{figure*}
	In \texttt{FIG.\,4b}, the temperature-dependent imaginary component of the linear susceptibility under an applied dc-field ($\chi_1^I$(T)) is depicted. The plot reveals an initial minor increase in the peak temperature (T$_{max}$) up to 100\,Oe of dc-field. Notably, the peaks of the graph for the dc-fields of 10\,Oe and 100\,Oe exhibit minimal variation, thus for these cases, the midpoint was taken as T$_{max}$, with corresponding error bars added to these data points, as illustrated in the inset of \texttt{FIG.\,4b}. Additionally, a decrease in the amplitude of the loss component is observed. Beyond 100\,Oe, or at higher dc-field amplitudes, T$_{max}$ decreases steadily, as indicated in the inset of \texttt{FIG.\,4b}, while the peak height of $\chi_1^I$ increases correspondingly.
	
	According to the Bean model \cite{r26}, the loss peak corresponds to the temperature at which the magnetic field penetrates the center of the superconducting specimen or the point at which the critical current density (J$_{C}$) of the superconductor diverges. At low dc-field amplitudes, the GdO component remains in a paramagnetic state, exhibiting no anomalies in the $\chi_1^I$(T) graph within this temperature range. However, at higher dc-field amplitudes, the spins of the paramagnetic component tend to align with the dc-field, resulting in a reduced response to the applied ac-field. Consequently, a loss component or a finite value of $\chi_1^I$(T) is observed in the ac-susceptibility measurement.

The observed decrease in the loss peak height up to a dc-field of 100 Oe is accompanied by a slight increase in the critical current density of B3. Consequently, a modest increase in T$_{\max}$ ($\sim$1,K) is noted. As the critical current density decreases beyond this field, the loss peak shifts toward lower temperatures. The unpaired spins of GdO may contribute to the rise in the magnitude of $\chi_1^I$. This effect arises because, as previously indicated, these spins become biased under the influence of the higher dc bias field and do not respond to the applied ac-field, thereby increasing the amplitude of $\chi_1^I$.

The vortex dipole interaction model can also explain the dc-field dependent behavior of T$_{min}$ and T$_{max}$.  Given that Gd$^{+3}$ ions possess a large spin magnetic moment ($\sim$7$\mu_{b}$)  even a small magnetic field can exert significant torque on them. At low temperatures, where the thermal fluctuation of Gd$^{+3}$ spins is suppressed, the interaction probability increases. When a dc-field is applied in this context, the thermal fluctuation of Gd$^{+3}$ spins reduces further at lower dc bias field values. Since the dc-field lacks sufficient energy to mobilize the vortices owing to its smaller amplitude, the interaction probability between the vortex and the magnetic dipole increases. This interaction enhances the vortex pinning strength, resulting in an increased amplitude of the critical current density, which in turn leads to an increase in and a decrease in the peak height of $\chi_1^I$.

Additionally, the increase in critical current density leads to a rise in the diamagnetic fraction of $\delta\chi_1^R$(T) and T$_{min}$  within the dc-field range of 0-100\,Oe. However, as the dc bias field exceeds 100\,Oe, the Lorentz force becomes strong enough to mobilize the vortices. This mobilization results in finite vortex motion, reducing the interaction probability between the vortex and the dipole. Consequently, T$_{\max}$ of $\chi_1^I$ shifts to a lower temperature, along with T$_{\min}$ of $\delta\chi_1^R$ and the diamagnetic fraction of $\delta\chi_1^R$(T) also decreases. Here the picture we want to convey is like,In type II superconductors, the interaction between vortices can mediate an indirect coupling between Gd$^{+3}$ spins. This vortex-mediated interaction leads to a ferromagnetic or ferrimagnetic ordering of the Gd$^{+3}$ spins. This dc-field dependent behavior clearly indicates that the anomalous electromagnetic shielding (ES) is predominantly due to the vortex-dipole interaction. Moreover, it is significant to note that the ES value at lower temperatures is relatively high (approximately $\sim$ 10$^{-5}$emu), suggesting that this phenomenon extends beyond the interface and impacts the bulk material as well.
	\begin{figure*}[t]
	\centering
	\includegraphics[width=16cm,height=12cm]{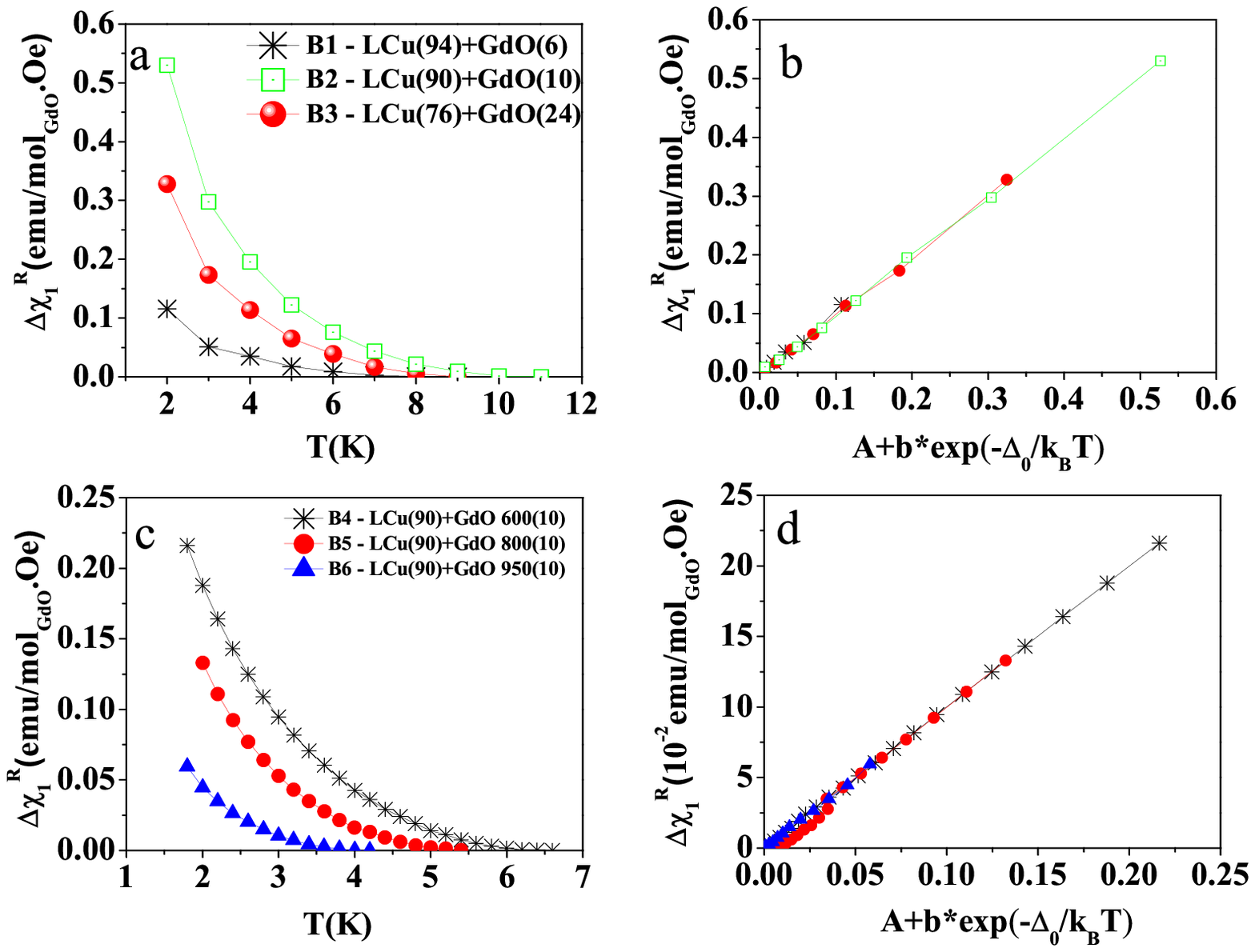}
	\vspace*{-2mm}
	\caption{\textit{\small{(Colour Online)(a) $\bigtriangleup\chi_1^R$ is plotted against temperature for the composite B1, B2 and B3.(b) $\bigtriangleup\chi_1^R$ is plotted against scaled temperature axis.(c) $\bigtriangleup\chi_1^R$ is plotted against temperature for the composite B4,B5 and B6.(d) $\bigtriangleup\chi_1^R$ is plotted against scated temperature axis.}}}
	\label{fig:fig5}
\end{figure*}
If the argument regarding excess susceptibility presented above is correct, then an internal field should have developed as a result of this ferrimagnetic (or ferromagnetic) type of modulated spin arrangement. This internal field should have been visible in the $\chi_2$ (second-order susceptibility) measurement without a superimposed DC field, and it can also be tuned by adjusting the vortex-dipole interaction (or by changing the DC field amplitude). The temperature-dependent plot of the second-order AC susceptibility for the composite B3 is shown in \texttt{FIG.\,5a} without a superimposed DC field ($\arrowvert\chi_2\arrowvert$, left-hand side Y-axis scale). It demonstrates that the anomaly in $\arrowvert\chi_2\arrowvert$ appears below 15\,K and a second anomaly is observed below 10\,K, where $\arrowvert\chi_2\arrowvert$ exhibits an almost divergent behavior. In any magnetic substance, $\arrowvert\chi_2\arrowvert$ appears in the presence of a symmetry-breaking field \cite{r36,r39,r42}. The anomaly of $\arrowvert\chi_2\arrowvert$ below 15\,K in zero DC field thus confirms that the ordering nature of the Gd$^{+3}$ spins is of the ferromagnetic (or ferrimagnetic) kind and is a signature of some kind of internal symmetry-breaking field.

Similar measurements were conducted for the parent compounds GdO and LCu. Because GdO is a paramagnetic material and exhibits a paramagnetic to antiferromagnetic (AFM) transition below 3 K, there is no anomaly in $\arrowvert\chi_2\arrowvert$ with zero superimposed DC field \cite{r44,r45,r46}. Since AFM has no internal field, $\arrowvert\chi_2\arrowvert$ remains zero down to the lowest measurable temperature. In the case of LCu, $\arrowvert\chi_2\arrowvert$ appears only in the presence of a superimposed DC field with the AC field or when the critical current density becomes a function of the applied DC field \cite{r27,r28,r29}. The temperature-dependent plot of $\arrowvert\chi_2\arrowvert$ at three superimposed DC fields (0, 50, and 100\,Oe) is shown in \texttt{FIG.\,5b}. It demonstrates that as the DC field amplitude increases, $\arrowvert\chi_2\arrowvert$ initially increases (up to 50\,Oe) and then decreases at higher values of the DC field (at and above 100\,Oe). These results reveal that the internal field amplitude increases up to a particular DC bias field value and decreases at higher DC field values, indicating a tunable behavior.

The modulated ferromagnetic (or ferrimagnetic) type interaction between Gd$^{+3}$ ions is therefore described by the DC field-dependent graph of $\arrowvert\chi_2\arrowvert$ and follows a similar DC field-dependent behavior as found in the case of $T_{min}$, $T_{max}$, $\chi_1^I(T)$, and $\delta\chi_1^R(T)$ (discussed previously). This description is also supported by the size effect analysis of GdO (\texttt{FIG.\,3a}), where it is seen that the proximity effect-induced ferromagnetic type of modulation between Gd$^{+3}$ spins increases as the AFM ordering tendency between the Gd$^{+3}$ spins decreases (i.e., with decreasing particle size shown in \texttt{FIG.\,3b}). This is why, as shown in \texttt{FIG.\,3a}, the low-temperature excess susceptibility in the $\delta\chi_1^R(T)$ graph and $T_{min}$ is observed at a higher temperature in the case of B4 compared to B5 and B6.
\begin{figure*}[t]
	\centering
	\includegraphics[width=17cm,height=14cm]{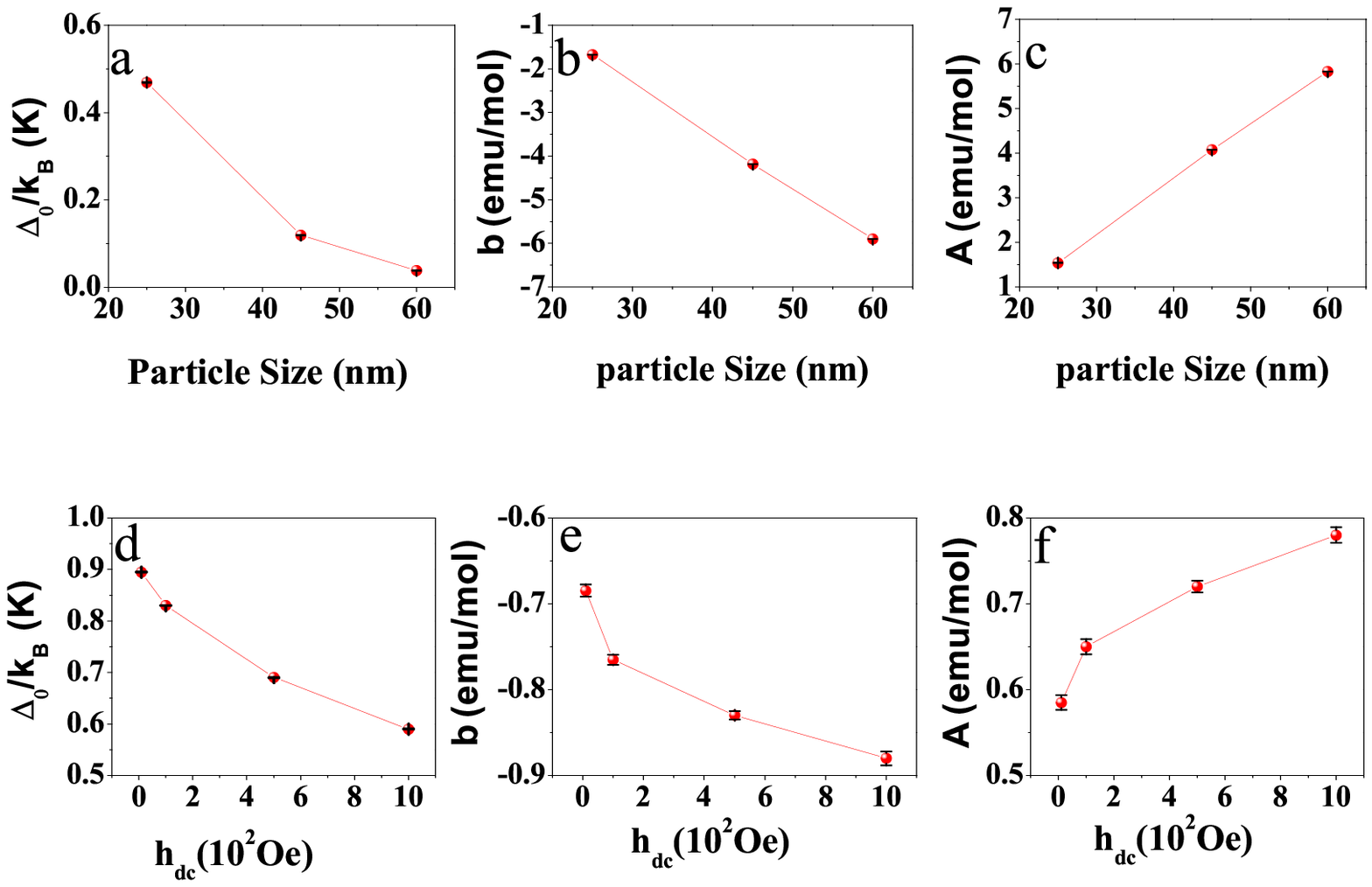}
	\vspace*{-2mm}
	\caption{\textit{\small{(Colour Online)(a) $T_0$ is plotted against the particle size.(b)  Parameter "b" is plotted against the particle size.(c) Parameter "A" is plotted against the particle size.(d) The exponential fitting of dc field superimposed susceptibility graph of composite B3.(e) and (f) shows the dc field dependent plot of T$_0$ and b. Inset shows the dc field dependent plot of parameter A.}}}
	\label{fig:fig6}
\end{figure*} 

\begin{table*}[t]
	\centering
	\caption{PS- Particle size.\label{lTABLE-2}}
	\begin{tabular}{cccccccccccccc }
		\toprule
		\multicolumn{1}{c}{Material} & \multicolumn{2}{c}{Gd$_{2}$O$_{3}$,(PS\,nm/$\mu$m)} &&
		\multicolumn{3}{c}{ h$_{dc}$(Oe)} &&&
		\multicolumn{4}{c}{		$\bigtriangleup$$\chi_1^R$\,(T)\,=\,A\,+\,b\,*exp(-$\frac{\bigtriangleup_0}{k{_B}T}$)}\\
		\cmidrule(rl){1-14}
		&&&&& &&&{A}&&&&{b} &{$\frac{\bigtriangleup_0}{k_B}$} \\
		\midrule
		B1-LCu(94)+GdO(6) &&Bulk( $\mu$m) &&0\,Oe&&& &1.364   &&&& -1.487 & 0.69 \\
		B2-LCu(90)+GdO(10)&&Bulk( $\mu$m) &&0\,Oe&&& & 1.44  &&&& -1.61&  1.09  \\
		B3-LCu(76)+GdO(24) &&Bulk( $\mu$m) &&0\,Oe&&& &0.357   &&&& -0.410 & 0.98  \\
		B3-LCu(76)+GdO(24) &&Bulk( $\mu$m) &&10\,Oe  &&& &  0.585   &&&& -0.68 & 0.89 \\
		B3-LCu(76)+GdO(24)&&Bulk( $\mu$m) &&100\,Oe&&& & 0.65   &&&& -0.76 & 0.83 \\
		B3-LCu(76)+GdO(24)&&Bulk( $\mu$m) &&500\,Oe&&& & 0.72   &&&& -0.83 & 0.69 \\
		B3-LCu(76)+GdO(24)&&Bulk( $\mu$m) &&1000\,Oe&&& & 0.78   &&&& -0.88 & 0.59 \\
		B4-LCu(90)+GdO600(10)&&25( nm) &&0\,Oe&&& & 1.54   &&&& -1.67 & 0.46 \\
		B5-LCu(90)+GdO800(10)&&45( nm) &&0\,Oe&&& & 4.07   &&&& -4.18 & 0.11 \\
		B6-LCu(90)+GdO950(10)&&60( nm) &&0\,Oe&&& & 5.83   &&&& -5.90 & 0.038 \\
		\bottomrule
	\end{tabular}
\end{table*}
The temperature-dependent graph of third-order susceptibility ($\arrowvert\chi_3\arrowvert(T)$) is also shown in \texttt{FIG.\,5a} (right-hand side Y-axis scale). It shows two anomalies: one at high temperature (25\,K - 12\,K) and another at low temperature (below 12\,K). Generally, $\chi_3$ appears due to the irreversibility of flux motion in a superconductor \cite{r26}. The high-temperature anomaly is due to irreversibility in flux motion inside the grain, and the lower temperature anomaly (below 10\,K) might be associated with the modulated ordering of Gd$^{+3}$ spins. Therefore, the fundamental susceptibility graph and the higher-order susceptibility graph represent the low-temperature anomalous excess susceptibility, which is a result of the magnetic interaction between Gd$^{+3}$ spins being modified by the vortex-dipole interaction, and this modulated interaction has either a ferromagnetic or a ferrimagnetic in nature.

	\maketitle\section{ Scaling analysis}
	The excess susceptibility ($\bigtriangleup\chi_1^R$\,(T)) at low temperature(i.e. below T$_{min}$) exhibits an exponential temperature dependent behavior. Which is an indication of some temperature dependent relaxation or opening of some kind of energy gap below T$_{min}$. The definition of $\bigtriangleup\chi_1^R$\,(T) is shown by \texttt{Eqn.\,1}. All these graphs can be scaled by a single universal equation (\texttt{Eqn.\,2}). \texttt{FIG.\,6a} shows the  plot of $\bigtriangleup\chi_1^R$\,(T) below $T_{min}$ of B1, B2 and B3 composite,  
	\begin{equation}
		\bigtriangleup\chi_1^R\,(T)\,=\,\delta\chi_1^R\,(T)\,-\,\delta\chi_1^{R}\,(T_{min})
	\end{equation} 
	
	\begin{equation}
		\bigtriangleup\chi_1^R\,(T)\,=\,A\,+\,b\,*exp(-\frac{\bigtriangleup_0}{k_{B}T})
	\end{equation}
	
	where A, b and $\bigtriangleup_0$ are constants.
	In \texttt{FIG.\,6b} the scaled graph of $\bigtriangleup\chi_1^R$\,(T) of the composites B1, B2, and B3 is displayed. Similarly $\bigtriangleup\chi_1^R$\,(T) plot of B4, B5 and B6 are shown in \texttt{FIG.\,6c}, and the corresponding scaled curves are shown in \texttt{FIG.\,6d}. and FIG. 6d illustrate that the nature of ES (or the change of the diamagnetic fraction ) can be described by a single, universal equation or curve, regardless of the effective
	interface between LCu and GdO. The respective values of the corresponding parameters used in \texttt{Eqn.\,2} (i.e.A,b, $\bigtriangleup_{0}$) of all the composites are given in \texttt{Table\,II}.  \texttt{FIG.\,7} illustrates how the associated parameters vary as a function of particle size and the DC field. The change of $\frac{\bigtriangleup_0}{k_B}$, b and A against particle size of GdO is shown in \texttt{FIG.\,7a}, \texttt{FIG.\,7b} and \texttt{FIG.\,7c}, respectively. Similarly, the variation of the corresponding parameters (i.e. $\frac{\bigtriangleup_0}{k_B}$, b and A) against dc field  are shown in \texttt{FIG.\,7d},\texttt{FIG.\,7e}  and \texttt{FIG.\,7f}, respectively.

	The amplitude of $\frac{\bigtriangleup_0}{k_B}$ is observed to decrease with increasing the particle size of GdO, $T_{min}$ also shows the similar behavior (Shown in the inset of \texttt{FIG.\,3a}). 
	The dc field dependent behaviour of $\frac{\bigtriangleup_0}{k_B}$ (shown in \texttt{FIG.\,7d}) below 100\,Oe is not so significant and above 100\,Oe the decrement is very significant, which almost mimic the dc field dependent behavior of  $T_{min}$.below 100 Oe is not so significant and above 100 Oe the decrement is very significant, which almost mimic the dc field dependent behavior of $T_{min}$. As previously stated, due to finite vortex motion, the strength of the vortex dipole interaction increases up to 100\,Oe of the dc bias field and subsequently decreases above h$_{dc}>$100\,Oe. The dc field superimposed $\arrowvert\chi_2\arrowvert$(T) measurement also shows an initial increase in ferromagnetic correlation at a lower dc bias field value (h$_{dc}<$100\,Oe), followed by a decrease in ferromagnetic (or ferrimagnetic) correlation at a larger dc field amplitude (i.e. h$_{dc}>$100\,Oe Oe). Additionally, as evidenced by the vortex dipole interaction and ferromagnetic (or ferrimagnetic) type correlation, the dc field dependent behaviour of $T_0$ and $T_{min}$ also exhibits similar behaviour.
		\begin{figure}[htbp]
		\centering
		\includegraphics[width=10cm,height=8cm]{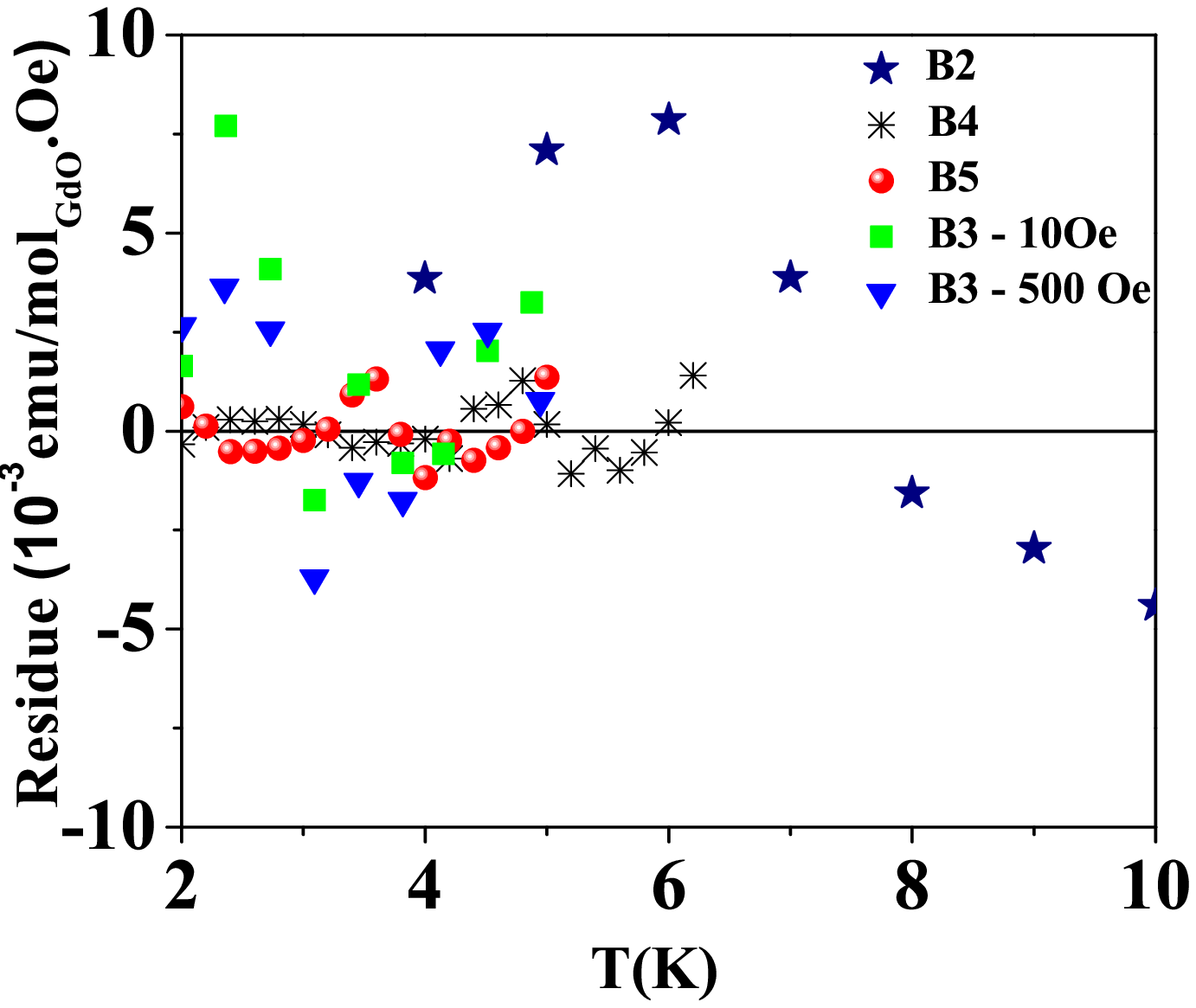}
		\vspace*{-2mm}
		\caption{\textit{\small{(Colour Online)Residue of the fitted graph of B2, B4, B5, B3 at different magnetic field are shown. }}}
		\label{fig:fig7}
	\end{figure} 
	
	The residue of all the fitted curve is shown in \texttt{FIG.\,8}. 
	The random ness of the residue across the zero point indicates the fitted equation used here is valid for the experimentally obtained graph. Therefore, Eqn.2 infer opening up of some type of energy gap much below the superconducting transition temperature and the second harmonic data indicates the presence of spontaneous magnetic moment in that particular energy state. Therefore, it can be inferred that the value of $\frac{\bigtriangleup_0}{k_B}$ and  $T_{min}$ is determined by the strength of the ferromagnetic or ferrimagnetic type of interaction between the Gd$^{+3}$ spins via vortex. According to \texttt{FIG.\,7b}, the parameter "b" decreases in a nonlinear manner as GdO particle size increases. however, it grows as the amplitude of the dc bias field grows (as illustrated in \texttt{FIG.\,7f}). The sharpness of the ES is defined here by the parameter "b," which also denotes the reduction of the diamagnetic fraction. As shown in \texttt{FIG.\,7b} and \texttt{FIG.\,7f}, the sharpness of the ES reduces with decreasing effective interface and increases with rising amplitude of dc bias field. The parameter "A" increases as the GdO particle size increases (as shown in \texttt{FIG.\,7c}), but it remains nearly same against the amplitude of the dc bias field (as seen in the inset of \texttt{FIG.\,7f}). This parameter can be related to the disorder at the bulk of GdO because the disorder decreases with increasing GdO particle size, which causes its effective negative value to decrease. However, it remains constant with dc field for a specific material (in this case, B3) because the disorder cannot be tuned by a magnetic field, so it shows almost constant value against dc magnetic field. Another important point is the tunability of the energy barrier with the magnetic field. The energy value of these states are almost $\sim$10$^{-6}$eV much lower than the energy gap of the superconductor LCu ($\sim$meV) and also spin polarized in nature, which we believe can be a suitable candidate for the YSR state. By examining the residue plotted in \texttt{FIG.\,8}, one may determine the correctness of the fitting.
	
	It displays the difference between the real experimentally
	obtained graph and the fitted graph. Here, the fitting has
	been done by taking into account the initial parameter
	$\bigtriangleup_0$=  k$_B$T$_{min}$ (k$_B$=8.61*10$^{-5}$ $\frac{eV}{K}$). The acquired values are then fed back into the equation a second time, iteratively fitting the graph.
	\vspace*{-5mm}
	\subsection{CONCLUSION}
	\vspace*{-5mm}
	Therefore, the excess susceptibility behavior at low
	temperatures and a finite value of  $\arrowvert\chi_2\arrowvert$(T) at zero bias dc field around the same temperatures represent a modulated ferromagnetic or ferrimagnetic interaction between Gd$^{+3}$ spins in GdO, with the modulation occurring as a result of vortex dipole interaction. The emergence of the spin triplet YSR state is indicated by this kind of activity. The Majorana zero mode is connected to the excitation of the YSR band, which is why the Majorana Fermion is being employed as a Q-bit in quantum computers. This	is the essential significance of the YSR state. This work thus points to a route for discovering YSR type Q-bits utilising high T$_C$ cuprate superconductors. Additional microscopic investigations, including as STM, STS, nutron scattering, etc., are necessary to demonstrate the existence of the spin triplet YSR state in these kinds of composite materials.
	\vspace*{-5mm}
	\section{Acknowledgment}
	\vspace*{-2mm}
	We thank Er. Kranti Kumar Sharma for fruitful discussions.
	
\end{document}